# Pitch Estimation by Denoising Preprocessor and Hybrid Estimation Model

Yu Cheng Hung
*Communication Engineering*
*National Taiwan University*
Taipei, Taiwan
r09942162@ntu.edu.tw

Ping-Hung Chen
*Communication Engineering*
*National Taiwan University*
Taipei, Taiwan
r07942055@ntu.edu.tw

Jian-Jiun Ding
*Communication Engineering*
*National Taiwan University*
Taipei, Taiwan
jjding@ntu.edu.tw

*Abstract—* Pitch estimation is to estimate the fundamental frequency and the midi number and plays a critical role in music signal analysis and vocal signal processing. In this work, we proposed a new architecture based on a learning-based enhancement preprocessor and a combination of several traditional and deep learning pitch estimation methods to achieve better pitch estimation performance in both noisy and clean scenarios. We test 17 different types of noise and 4 $SNR_{db}$ noise levels. The results show that the proposed pitch estimation can perform better in both noisy and clean scenarios with short response time.

## I. Introduction

In general, pitch estimation follows the process of short time analysis or note duration for input audio as follows. 1) Get the note duration from onset detection. 2) Compute the pitch of each frame by applying different methods. 3) Get several high peaks that contain fundamental frequency ($f_0$) and its harmonics and a small peak before $f_0$ which is sub-harmonics.

The harmonics and sub-harmonics are represented as

$$n \times f \text{ and } \frac{1}{n}^{th} \times f \qquad (1)$$

where $f$ is $f_0$ and $n$ is an integer. Under clean signal scenarios, most time domain methods like the auto-correlation function (ACF) and normalized dynamic spectral features (NDSFs) [1], and frequency domain methods like the cepstrum [6] and the harmonic product spectrum (HPS) [2] work well.

The main concept of the NSDF and the ACF [1] is the inner product of the overlap part between the original frame $x(t), t = 0, \dots n - 1$ and the shift of frame $x(t)$ with $\tau$ delayed in terms of $x(t - \tau)$. $\tau$ is the time lag of the original signal.

The HPS [2] is to compute the product of the spectrum of audio and its compressed spectrum of audio. The property of HPS is that each compressed spectrum has a peak around its $f_0$, then the product of all spectrum makes the $f_0$ even higher value. The short time Fourier transform (STFT) [3] is common in the audio and image processing realm because STFT outputs a spectrogram as in Fig. 1. If we sum up the energy along the frequency axis from the transposed spectrogram, we can obtain the energy distribution of a single note in frequency. The max likelihood (ML) method is to use an impulse function with $f_0$ and its 4 octaves. Then convolve the impulse function with the spectrum of a note and sum over the value to get the pitch $f_0$ of a note.

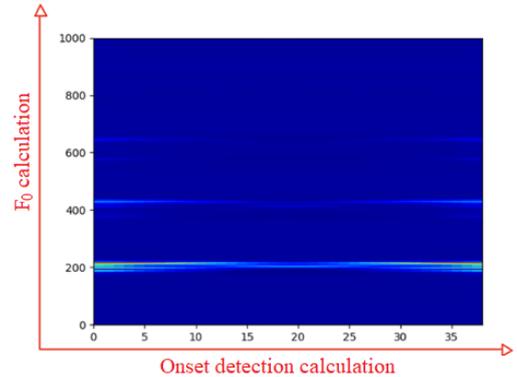

Fig. 1. Spectrogram for pitch estimation and onset detection

TABLE I. Comparison Of Different Methods

| Methods | Formula |
|---|---|
| HPS | $HPS(f) = \prod_{i=1}^{n}\|X(nf)\|, \quad f_{f0} = \max_f(HPS(f))$ |
| STFT | $E(f) = \sum_t X(t,f), \quad f_{f0} = \max_f(E(f))$ |
| Max likelihood | $f_{f0} = \max_k \sum_f X(f)\delta(f - nk), n \in \mathbb{N}$ |
| Cepstrum | $C(n) = real(\int_{-\frac{1}{2}}^{\frac{1}{2}} \log\|X(f)\| e^{j2\pi fn} df), \quad f_{f0} = \frac{fs}{\max Cep(n)}$ |
| SRH | $SRH[f] = E[f] + \sum_{k=2}^{N_{harm}}[E[kf] - E[(k - \frac{1}{2})f]]$ $f_{f0} = \max_f(SRH[f])$ |

The summation of residual harmonics (SRH) [4] is based on speech signals. In speech analysis, speech can be deconvolved into excitation signal and formant structure. It uses linear prediction analysis to estimate excitation. The deep learning (DL) method Crepe method [5] is data-driven and the training is computation cost, we only use the pre-trained model to improve the pitch analysis. The overview of pitch method research is in Table I.

## II. Proposed Pitch Estimation

In this work, we proposed a robust and reasonable response time with an ensemble of several traditional methods and a deep learning method in the low SNR scenario. Here we assume the onset locations are already known and test different methods for clean audio, different noisy audio, and response time.

Each method has its advantage and disadvantage. For the ML and the STFT methods, they tend to be influenced by noise within low-frequency noise like in offices, train stations, and ventilation. Using the concept of machine learning ensemble, we combine multiple methods that have small pitch error in both clean, noisy audio and short response time within 2s to get a robust $f_0$ estimation.

TABLE II. DIFFERENT SETTING FOR DIFFERENT PITCH ESTIMATION

| Method | Min frequency | Max frequency | Harmonics number |
|---|---|---|---|
| HPS | 0 | Sample length / 2 | 3 |
| STFT | 20 | 1000 | 4 |
| Max Likelihood | 20 | 800 | 5 |
| SRH | 80 | 500 | 5 |
| CREPE | 33 | 3951 | - |

Our idea is to find the property below and apply the additional algorithms to these methods. 1. Small pitch error in both clean and noisy audio. 2. noisy audio includes SNR -5, 0, 10, and 20 with 17 different noises. SNR 30 is almost near ground truth while -10 is not hearable. 3. Short response time within 2s per song average from 10 ~ 20s songs. We take the median of HPS, STFT, ML, SRH, and Crepe with tiny pre-trained model. We set the different frequency ranges for different methods to avoid high-frequency noise beyond 1000 Hz as Table II. The different frequency ranges can be further researched in future work. In our research, we just empirically set the minimum frequency, maximum frequency, and the harmonics numbers and then ensemble all the methods.

## VI. SIMULATION RESULTS

For noise data, we use the UBC noise data set containing 17 different noise which covers most scenarios for QBH usage as one can imagine and the noise level is [-5, 0, 10, 20]. The song data from our lab contains about 209 songs. Each song has 68 scenarios to consider. Table III is the overview of Tables IV and V which is the average of noises 1 to 17.

Our proposal outperforms the other method in 14 scenarios and achieves equivalent in 3 scenarios. The pitch estimation measurement is to calculate the pitch difference as

$$error = \frac{1}{N}\sum_{n=1}^{N}\sqrt{|f_{method(n)} - f_{groundtruth(n)}|}, \quad (2)$$

where $n$ is the total number note of a song.

In our experiments, we make an overview comparison for all methods described in Tables IV and V followed by the error function in (2) with 17 different noises including 1. White, 2. Babble, 3. Insect, 4. Surf, 5. Subway, 6. Campus, 7. Ventilation, 8. Car, 9. Train, 10. Conservator, 11. Exhibition, 12. Gaussian, 13. Wilderness, 14. Restaurant, 15. Airport, 16. Street, 17. Office. As in Tables IV and V, our proposal outperforms the other methods in 13 scenarios but achieves equivalently as the other methods in 4 scenarios. The measurement value is an average error with noise level [-5 ~ 20].

## V. CONCLUSION

In this paper, we provide new insight with a combination of traditional and deep learning methods. By combing different methods with the median, we can avoid most of the noise affecting pitch estimation and get a robust pitch estimation. In the future, we can further research the relationship between different frequency settings and different methods.

TABLE III. DIFFERENT MEASUREMENTS FOR DIFFERENT ALGORITHMS.

| | SRH | ML | HPS | STFT | Crepe | Ours |
|---|---|---|---|---|---|---|
| Clean audio error | 1.6 | 2.34 | 2.05 | 2.38 | 2.57 | **1.48** |
| Noisy audio error | 2.59 | 3.6 | 3.12 | 4.07 | 3.13 | **2.46** |

TABLE IV. PITCH ESTIMATION ERRORS FOR NOISE 1~9 WITH $SNR_{db}$ = [-5, 0, 10, 20].

| Noise \ Method | 1 | 2 | 3 | 4 | 5 | 6 | 7 | 8 | 9 |
|---|---|---|---|---|---|---|---|---|---|
| SRH | 2.06 | 2.93 | 1.79 | 3.03 | 2.61 | 2.49 | 3.26 | 2.83 | 2.74 |
| YIN | 8.06 | 7.79 | 5.32 | 8.05 | 7.81 | 8.06 | 8.03 | 8.15 | 8.41 |
| NSDF | 2.67 | 3.42 | 4.31 | 4.12 | 3.23 | 6.76 | 5.57 | 6.58 | 7.95 |
| HPS | 2.09 | 2.69 | 2.06 | 2.84 | 2.08 | 4.35 | 4.16 | 3.82 | 5.09 |
| ML | 2.51 | 3.29 | 2.86 | 3.76 | 2.83 | 4.87 | 4.49 | 4.32 | 4.99 |
| STFT | 2.63 | 3.45 | 2.39 | 3.45 | 2.90 | 6.15 | 5.11 | 4.90 | 6.43 |
| CREPE | 2.85 | 3.34 | 2.89 | 3.60 | 3.05 | 2.85 | 3.16 | 3.07 | 2.99 |
| Cepstrum | 12.95 | 13.56 | 13.50 | 13.43 | 13.64 | 13.70 | 13.55 | 13.33 | 13.51 |
| **Proposed** | **1.59** | **2.19** | **1.50** | **2.34** | **1.69** | **3.68** | **3.15** | **3.15** | **4.07** |

TABLE V. PITCH ESTIMATION ERRORS FOR NOISE 10~17 WITH $SNR_{db}$ = [-5, 0, 10, 20].

| Noise \ Method | 10 | 11 | 12 | 13 | 14 | 15 | 16 | 17 |
|---|---|---|---|---|---|---|---|---|
| SRH | 2.48 | 2.67 | 2.04 | 2.97 | 2.96 | 3.45 | 3.00 | 2.28 |
| YIN | 8.06 | 7.95 | 7.49 | 8.02 | 7.78 | 8.10 | 7.69 | 7.89 |
| NSDF | 3.83 | 3.40 | 2.82 | 4.25 | 3.88 | 5.80 | 4.48 | 8.88 |
| HPS | 2.53 | 2.08 | 2.09 | 2.87 | 2.38 | 3.82 | 2.83 | 5.34 |
| ML | 3.08 | 2.52 | 2.50 | 3.75 | 3.07 | 3.93 | 3.82 | 4.55 |
| STFT | 3.48 | 3.09 | 2.68 | 3.46 | 3.22 | 4.79 | 4.39 | 6.69 |
| CREPE | 3.13 | 2.88 | 2.84 | 3.41 | 3.32 | 3.41 | 3.59 | 2.93 |
| Cepstrum | 13.18 | 13.37 | 13.15 | 13.61 | 13.57 | 13.57 | 13.52 | 13.65 |
| **Proposed** | **1.94** | **1.62** | **1.57** | **2.35** | **2.01** | **2.86** | **2.44** | **3.75** |